\documentclass[aps,physrev,twocolumn,superscriptaddress]{revtex4-2}

\usepackage{graphicx}% Include figure files
\usepackage{dcolumn}% Align table columns on decimal point
\usepackage{bm}% bold math
\usepackage[utf8]{inputenc}
\usepackage[T1]{fontenc}
\usepackage{mathptmx}
\usepackage{etoolbox}

\begin{document}

\title[A continuously-cooled $^3$He/$^4$He phase separation refrigerator]{A continuously-cooled $^3$He/$^4$He phase separation refrigerator}

\author{P.H. Kim}
 \affiliation{Zero Point Cryogenics, Edmonton, AB T6E 5V8, Canada}
 \author{M. Hirschel}
 \affiliation{Zero Point Cryogenics, Edmonton, AB T6E 5V8, Canada}
\author{J. Suranyi}%
 \affiliation{Zero Point Cryogenics, Edmonton, AB T6E 5V8, Canada}
\author{J.P. Davis}
 \affiliation{Zero Point Cryogenics, Edmonton, AB T6E 5V8, Canada}

\date{\today}% It is always \today, today,

\begin{abstract}
We present a novel cooling method that uses the phase separation and evaporative cooling of $^3$He to reach and continuously maintain sub-kelvin temperatures. While less complex than a dilution refrigerator, the system performs similarly to a continuous $^3$He cryostat but with a simpler design, a more efficient cooldown process, and  a significantly smaller $^3$He requirement. Our prototype demonstrated a base temperature of 585~mK and 3~mW cooling power at 700~mK using just two gaseous liters of $^3$He.  Lower temperatures could be expected in systems with improved heat exchangers and pumping efficiency.
\end{abstract}

\maketitle

\section{\label{sec:intro}Introduction}

Many aspects of the physical sciences must be studied at low temperatures, frequently well below 1~K, including the study of quantum materials and quantum circuits.  For example, the quantum bits, or qubits, that make up modern quantum computers must be operated in a regime at or near the ground state occupation of their respective quantum harmonic oscillator\cite{ladd2010}. Furthermore, superconducting qubits need to be cooled to a regime without circuit resistance---taking advantage of superconductivity, itself a quantum phenomenon. Beyond that, a variety of other quantum computing architectures require low temperatures, reducing loss to phonons in color centers or taking advantage of superconductivity in transition edge sensors for photonic computing.

With rapid advances in quantum research and technology as well as the growth of the quantum computing industry, there is an increasing demand for simple, user-friendly, and affordable solutions that provide continuous cooling at sub-kelvin temperatures. While the cryogen-free (“dry”) dilution refrigerator is currently the most popular platform, its low base temperature, on the order of 10~mK, exceeds the requirements for an increasing number of systems, particularly as qubit frequencies are raised to reduce the thermal photon occupation. Instead, less complex systems promise affordable solutions to provide sufficient cooling power at a few hundred millikelvin temperatures\cite{anferov2024}. 

In Sec.~\ref{sec:cooling_tech}, we review cooling techniques to obtain temperatures below 4K---$^4$He and $^3$He cryostats, as well as dilution refrigerators and adiabatic demagnetization refrigerators (in order of decreasing base temperature). While continuous $^4$He systems have become quite prevalent, as have continuous, dry dilution refrigerators, continuous $^3$He systems are quite rare.  This may be because of the added complexity and cost of currently realized incarnations of these systems.  

We find three examples of such cryostats.  Firstly,  Burton et al.~described a continuous, dry, $^3$He refrigerator in 2011 that condenses $^3$He into a pot using a separate continuous $^4$He system\cite{burton2011}.  Hence, two gas handling systems and associated pumping systems must be used, one for the $^4$He and one for the $^3$He, although we note in this example that the authors used a simple gas handling system for the $^4$He circuit that vented the helium to the atmosphere.  They reached a base temperature of 0.36~K, with a cooling power of 1.75~mW at 0.5~K.  The second reference is in the chapter of a book by Zhao and Wang\cite{zhao2019}, discussing the concept of a dry, continuous $^3$He refrigerator that used separate gas handling circuits for $^3$He and $^4$He.  Finally, we note that a system similar to the latter has been commercialized by ICEOxford and is available for purchase\cite{dryice300mK}.  

At Zero Point Cryogenics, we have discovered a novel cooling method and developed a dry, continuous, $^3$He/$^4$He refrigerator that is much simpler to use than the above three references. Additionally, it can operate with significantly smaller amounts of $^3$He, which is critically important due to the scarcity and expensiveness of this isotope.  The working principle relies on the phase separation of $^3$He and $^4$He, but evaporative cooling instead of a dilution process as used in a dilution refrigerator. Sec.~\ref{sec:system} explains the system design, before Sec.~\ref{sec:results} summarizes a characterization of our prototype. Finally, we conclude our study in Sec.~\ref{sec:conclusion}.

\section{\label{sec:cooling_tech}Cooling Techniques Below 4~K}

\begin{figure}[b]
\includegraphics[width=\columnwidth]{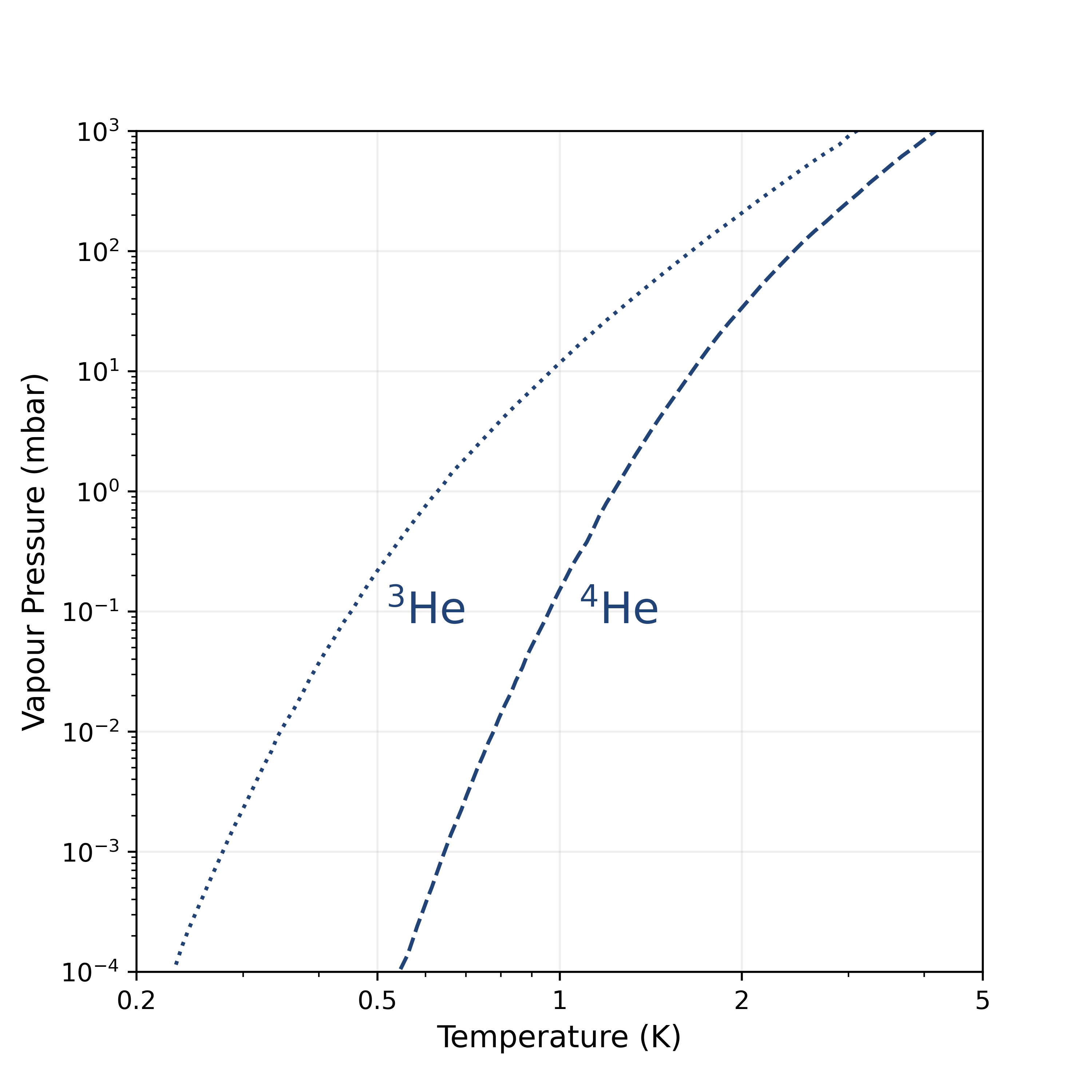}
\caption{\label{fig:vapor_pressure}The saturated vapor pressure (SVP) of liquid $^3$He and $^4$He as a function of temperature, decreasing exponentially upon cooling, but with $^3$He featuring a larger SVP at all temperatures\cite{pobell2007}.}
\end{figure}

Methods for achieving low temperatures are of great importance to both science and technology.  Beyond cooling to 77~K or 4.2~K using liquid nitrogen or liquid helium at standard pressure, respectively, one must harness the tools of thermodynamics, such as reducing the pressure by pumping on these liquid cryogens to achieve evaporative cooling.  By pumping on liquid $^4$He, one can cool below 1~K, typically limited to approximately 800~mK by the exponentially vanishing vapor pressure of liquid helium at these temperatures\cite{delong1971,wang2006}, see Fig.~\ref{fig:vapor_pressure}.  In a traditional “wet” fridge, which uses liquid $^4$He as a cryogenic bath, the liquid helium is let into a “pot” -- a small volume separated from the main bath by a fixed or variable flow impedance. To facilitate evaporative cooling, the helium is pumped out of the pot with an external pump before it is vented to the room or recovered and re-condensed by a separate cryogenic system.  Cooling in this way is only continuous so much as the bath of liquid helium is periodically refilled.  

To achieve even lower temperatures, one can switch to the only other stable isotope of helium, $^3$He.  It is exceedingly rare in naturally occurring helium (in the range of 1~ppm, depending on its origin) but is produced in nuclear reactions. $^3$He, being lighter than the common isotope, does not liquefy until 3.2~K at standard pressure and has a higher vapor pressure compared to $^4$He at an identical temperature.  Therefore, $^3$He can be used for evaporative cooling to lower temperatures, practically limited to approximately 300~mK\cite{mate1965,batey1998}.  

Because of the scarcity---and expensiveness---of $^3$He, one should never vent  $^3$He to the atmosphere.  Thus, pumped $^3$He refrigerators are typically operated in a closed-cycle and one-shot method\cite{mate1965}.  For this, a pumped $^4$He pot operating at around 1.5~K serves to condense a small volume of $^3$He into a separate, sealed volume in thermal contact with the lowest temperature stage and experimental payload.  This $^3$He pot is connected to a charcoal adsorption pump (or “sorb”).  When cooled below 10~K, the large surface area of the contained charcoal adsorbs the $^3$He, pumping and collecting it from the pot to obtain the desired evaporative cooling.  This is by its nature a “one-shot” system---when all of the liquid $^3$He has been pumped from the pot, it will no longer have any cooling power.  At this point, the sorb has to be heated with an externally controlled heater to 40~K or more, which desorbs and pushes the helium toward the lower temperature stages for recondensation into the $^3$He pot, such that the cycle can restart.  This causes a low-duty cycle, interrupting tests being performed on the cryogenic system.  Note, in all known examples of $^3$He systems, the $^3$He and $^4$He pumping lines (or pumping “circuit”) are kept entirely separate, such that the two isotopes do not mix.

\begin{figure}[t]
\includegraphics[width=\columnwidth]{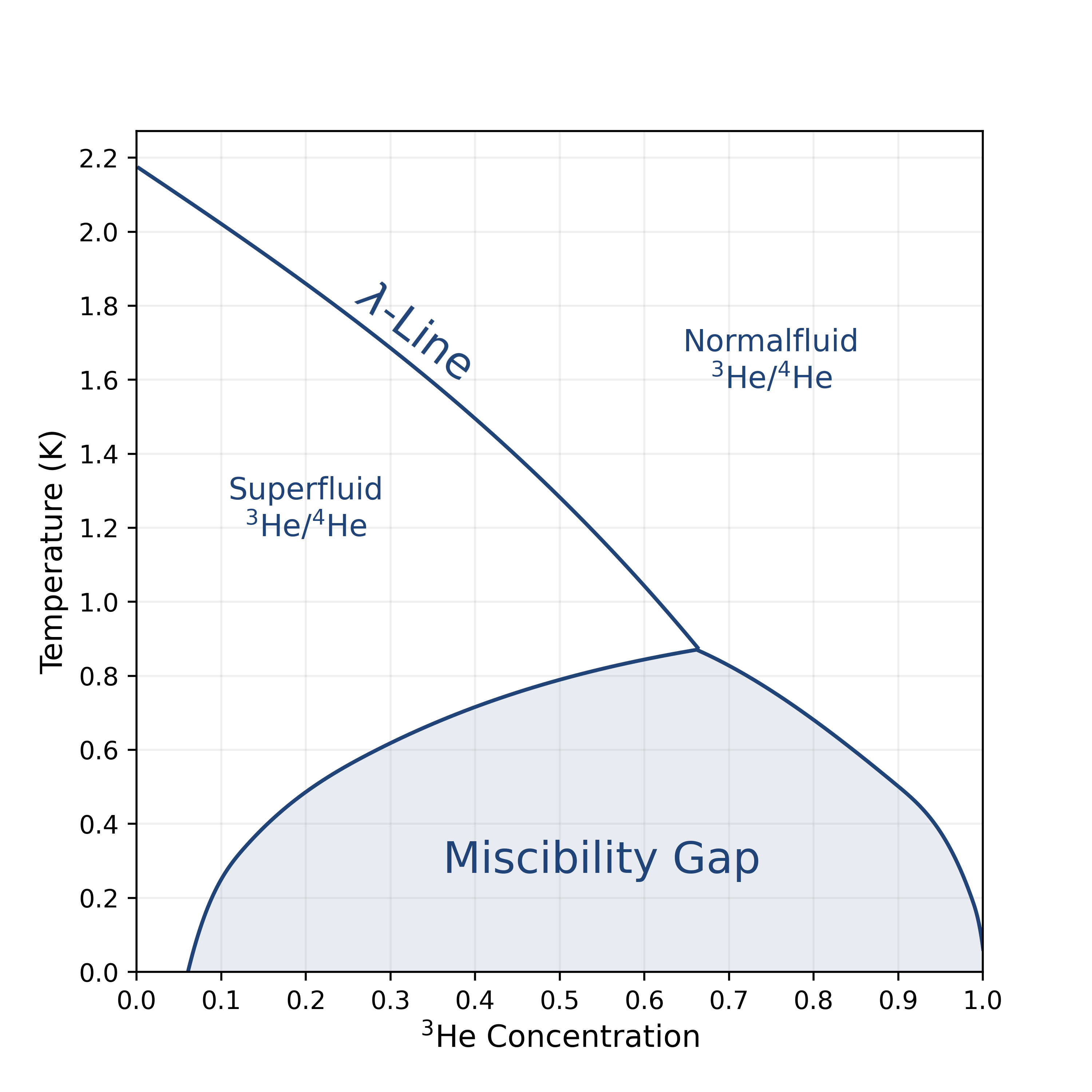}
\caption{\label{fig:phase_separation}Phase diagram of liquid $^3$He/$^4$He mixtures at saturated vapor pressure, with separation into a $^3$He-rich and a $^4$He-rich phase occurring naturally below 872~mK\cite{pobell2007}.}
\end{figure}

To achieve temperatures below that possible using pumped $^4$He or pumped $^3$He, one must either use a non-helium-based technique called adiabatic demagnetization \cite{kurti1956,kurti1960}, which cools using the temperature-dependent entropy of a paramagnetic system in a magnetic field\cite{vilches1966}, or by mixing the two isotopes of helium to make a dilution refrigerator\cite{london1962,frossati1978, cousins1999}. Adiabatic demagnetization is inherently one-shot and can result in unwanted stray and changing magnetic fields, in addition to having low cooling powers and, hence, being susceptible to external heat leaks.  Dilution refrigerators, on the other hand, can provide continuous cooling down to as low as 2~mK\cite{frossati1978}. Such systems take advantage of the phase separation that naturally occurs between the two isotopes of helium below 872~mK at a concentration of 67.3\% $^3$He in $^4$He\cite{goellner1973, wheatley1968}, as shown in Fig.~\ref{fig:phase_separation}.  Once phase-separated, the $^3$He can be pumped from the dilute concentration side, which results in an osmotic pressure that continuously drives $^3$He across the phase boundary from the concentrated to the dilute side, leading to continuous cooling power to extremely low temperatures.  The new system described below will also take advantage of the phase separation that occurs between the two stable isotopes of helium but is not a dilution refrigerator in that its cooling power does not come from this process of dilution. 

With the advent of “cryogen-free” technology, namely Gifford-McMahon (GM) coolers and pulse tube (PT) coolers, new “dry” (that is, without a bath of liquid cryogens) incarnations of the $^4$He and $^3$He cryostat\cite{paine2013, demann2016, klemencic2016}, as well as the dilution refrigerator have been developed\cite{uhlig1994,uhlig2002, zheng2019, uhlig2017, dewit2019, lounasmaa1979}.  Generally, instead of a pumped $^4$He pot, these systems will take advantage of a Joule-Thomson (JT) expansion process for cooling, which was absent from the wet versions.  In the simplest continuous “1~K” or “1.5~K” system\cite{wang2014}, $^4$He can be condensed by the PT cooler, operating below 4.2~K, generally at an elevated pressure brought about by the output port of a mechanical pump (optionally followed by a compressor).  The helium expands through an impedance or “Joule-Thomson plug,” resulting in cooling\cite{wilkes1972}.  The opposite side of the impedance is kept at a low pressure by a mechanical pump.  In a continuous system, the same helium that is pumped out of one side of the impedance can be re-condensed by the GM or PT cooler and once again expanded.  After a few cycles of this recirculation process, the cold side of the impedance can collect liquid $^4$He.  As a result, the cooling power is then exactly as in the wet $^4$He system, determined by the vapor pressure of $^4$He.  Similarly, cryogen-free dilution refrigerators also use a JT impedance instead of a $^4$He pot for liquefying the incoming $^3$He / $^4$He mixture.

\section{\label{sec:system}System Description}

\begin{figure}[t]
\includegraphics[width=\columnwidth]{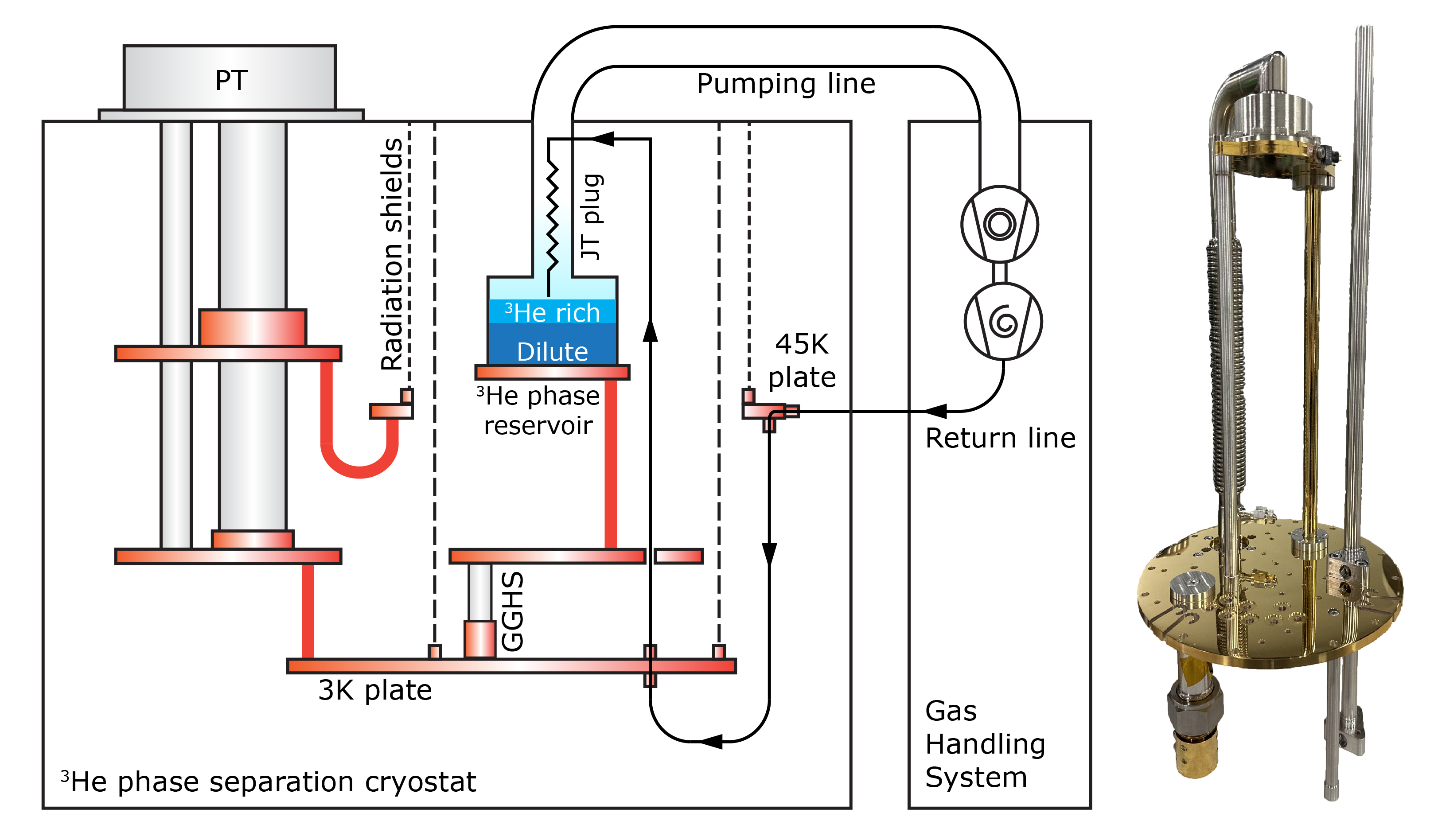}[t]
\caption{\label{fig:schematic} A schematic and a photo of the $^3$He/$^4$He phase separation refrigerator is shown in (a). The ZPC model-I cryostat consists of an outer chamber that holds a high vacuum and contains two radiation shields, kept at 45~K and 3~K through thermalization to the first and second stage of a pulse tube cryocooler, respectively. For initial pre-cooling, the phase separation reservoir is thermalized using a gas-gap heat switch linked to the 3~K stage. Subsequently, circulation of the $^3$He/$^4$He mixture through the JT plug leads to further cooling and then phase separation. The mixture is continuously pumped and returned to the cryostat through a gas handling system at room temperature.}
\end{figure}

Our system is a modified Zero Point Cryogenics Model-I cryostat, schematically shown in Fig.~\ref{fig:schematic}a.  It combines the $^3$He and $^4$He pumping circuits into a single pumping system by mixing $^3$He and $^4$He at a concentration of approximately 25\% $^3$He (although concentrations as low as 10\% are possible, there is no upper concentration limit). Note that it can be easily adapted to other fridge layouts. The cryostat is pre-cooled to around 3~K through a pulse tube cryocooler and gas gap heat switch thermally linking the low-temperature stages.  Once pre-cooled, the $^3$He/$^4$He mixture is injected into the system through the return line, where it undergoes Joule-Thomson expansion after a flow impedance, the value of which is in the typical range for a dilution refrigerator, $10^{10}-10^{12} \ \mathrm{cm}^{-3}$.  This can be aided by a compressor that increases the pressure in the return line, but is not strictly necessary, especially when operating with a low concentration of $^3$He in $^4$He.  With its higher liquefaction temperature, the $^4$He in the mixture condenses sooner and accelerates the Joule-Thomson expansion cooling, which is significantly more powerful when the system is cold enough for condensation.  Quickly, the system cools to the cryocooler's base temperature, and the heat switch is opened.  The system has a reservoir, essentially identical to the “still” of a dilution refrigerator, that accumulates liquid helium after JT expansion.  While in the early stages, all of the cooling is performed by the JT expansion, the cryostat eventually cools predominantly through evaporative cooling of the liquid helium.

As the $^3$He/$^4$He mixture is pumped from the reservoir, it will reach a temperature between 500 to 870~mK at which phase separation occurs, depending on the initial concentration of $^3$He.  Subsequently, the concentrated side---that is the $^3$He-rich side of the phase separation---will float on top of the dilute (or $^4$He-rich) side due to the smaller mass density of $^3$He.  This is the key to the operation of this system at lower temperatures than would be possible without phase separation.  The evaporative cooling in the reservoir occurs primarily from the floating $^3$He.  Its higher vapor pressure allows this system to operate similarly to a $^3$He refrigerator, with nearly identical base temperatures and cooling powers---limited by the finite concentration of $^4$He in the $^3$He---but requiring a much smaller amount of $^3$He.  Note that the finite solubility of $^4$He in the $^3$He will slightly raise the temperature of operation due to the evaporation of some $^4$He instead of $^3$He. 

\section{\label{sec:results}Results and discussion}

\begin{figure}[t]
\includegraphics[width=\columnwidth]{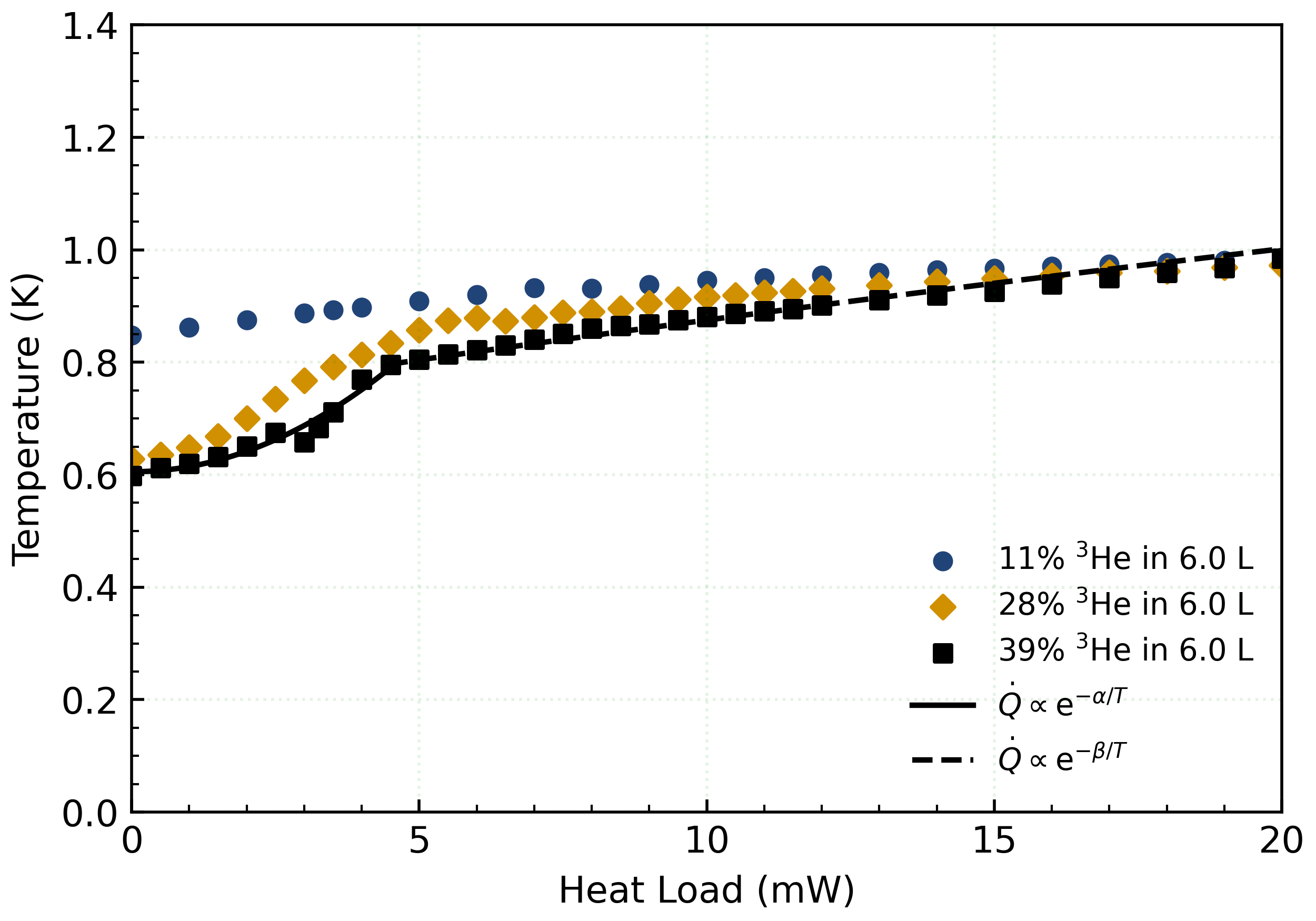}
\caption{\label{fig:cooling_power}Base temperature of the continuous $^3$He/$^4$He refrigerator as a function of the applied heat load, at $^3$He concentrations of 11\% (blue circles), 28\% (gold diamond), and 39\% (black square), respectively. The black lines (solid and dashed) show exponential fits to the 39\% concentration results, as expected when cooling through the evaporation of $^3$He and $^4$He, respectively.  It is noteworthy that 28\% and 39\% were sufficient concentrations to undergo phase separation and operate as described in the manuscript.  On the other hand,  the 11\% data set did not phase separate and, therefore, simply operated as a combined $^3$He and $^4$He evaporative cooler, without the advantages of the mechanism described here.}
\end{figure}

Fig.~\ref{fig:cooling_power} shows data from one incarnation of this system, with a concentration of 39\% $^3$He in $^4$He (corresponding to only 2.3 liters of $^3$He within a total mix volume of 6.0 liters).  The system achieved a base temperature of 585~mK, limited on the one hand by the heat load from the returning $^3$He and, on the other hand, by limitations in the pumping impedance.  In practice, more efficient heat exchangers at this stage will allow for lower base temperatures.  To characterize the cooling performance of the cryostat, an increasing amount of heat is applied to the base plate, which is thermally connected to the reservoir. Until a heat load of 3~mW is reached, the temperature remains below 0.7~K.  Above this heat load, the $^3$He has mostly been evaporated and no concentrated film floating on the dilute $^4$He remains.  Above approximately 5~mW of applied heat, the temperature rises gradually and the system operates essentially as a continuous $^4$He system.  Finally, at larger heat loads, the only cooling power in the system is provided by the JT expansion of the returning $^3$He/$^4$He mixture. 

Efficient heat exchange between the incoming and the outgoing $^3$He/$^4$He mixture is necessary to achieve the lowest temperature possible.  Additionally, to ensure that primarily $^3$He instead of $^4$He is pumped from the system, similar procedures to the ones used in a dilution refrigerator can be applied, such as a superfluid $^4$He film burner, knife-edge, or an orifice\cite{kirk1974}. We note, however, that these will be less useful in this system than for the still of a dilution refrigerator because the reservoir will generally be operated at lower temperatures.  Instead of deliberately raising its temperature with a heater to promote the enhanced circulation of $^3$He, one will achieve the lowest temperatures by pumping as powerfully as possible on the reservoir.  Thus, the use of an orifice, as was done in this prototype, is likely to be counterproductive.  It is also important to note that to pump the helium as efficiently as possible, the pumping line should be designed with a sufficiently small impedance at all temperatures so as not to limit the pumping capacity. Moreover, pumps efficient at pumping helium---both mechanical, such as scroll or roots pumps, and booster pumps, such as turbo pumps---should be used. 

Finally, we comment on the amount of $^3$He used in this system.  When operating in the phase separation regime, adding additional $^3$He to the pumping circuit does not significantly affect the base temperature, as it is already concentrated $^3$He being pumped from the reservoir.  Therefore, if low cooling powers are sufficient, it is advisable to use the minimum concentration of $^3$He in $^4$He to achieve phase separation at the temperatures achievable by pumping on the mixture---practically, we find that 25 to 30\% initial $^3$He concentration is nearly ideal. With the initial concentration of 11\% $^3$He, our system could not reach the miscibility gap through evaporative cooling, as shown by the data in Fig.~\ref{fig:cooling_power}. If one requires a higher cooling power at the lowest temperatures, additional $^3$He can be added to the system.  This leads to a larger reservoir of concentrated $^3$He in the reservoir, increasing the evaporative cooling capacity to balance the incoming heat load from an experiment. While the smallest suitable $^3$He concentration comes with the lowest overall cost of the system, there is no conceptual maximum, with 100\% concentration being the limit of a continuous $^3$He refrigerator.\\

\section{\label{sec:conclusion}Conclusion}

In conclusion, we have invented a new type of refrigerator that lies between a continuous liquid helium cryostat and a dilution refrigerator.  It has a similar base temperature and cooling power as a $^3$He refrigerator, in our first demonstration 585~mK and 3~mW at 700~mK, respectively. An incarnation with more efficient heat exchangers and improved pumping performance may reach temperatures as low as 300~mK\cite{mate1965,batey1998}. 

In contrast to a traditional $^3$He cryostat, however, our system is continuous, can easily be made cryogen-free, and takes advantage of the phase separation of $^3$He and $^4$He to provide a concentrated $^3$He surface for pumping. This minimizes the complexity of the pumping system and the amount of $^3$He needed.  The latter is particularly important given the limited availability and high price of $^3$He.  We expect this system to find wide adoption in the field of quantum technologies, where the reduced complexity and price while enhancing the ease of use will make this an attractive alternative to a dilution refrigerator.

\section*{Data Availability}
The data that supports the findings of this study are available within the article. 
\nocite{*}
\bibliography{references}% Produces the bibliography via BibTeX.

\end{document}